\documentclass{article}
\newcommand{\p}[1]{(\ref{#1})}
\newcommand{\cZ}{{\cal Z}}
\newcommand{\bz}{{\bar z}}
\newcommand{\bp}{{\bar p}}
\newcommand{\hp}{{\hat p}}
\newcommand{\bhp}{{\hat{ \overline p}}}
\newcommand{\cbZ}{\overline{\cal Z}}
\newcommand{\cF}{{\cal F}}
\newcommand{\cK}{{\cal K}}
\newcommand{\cV}{{\cal V}}
\newcommand{\cbF}{\overline{\cal F}}
\newcommand{\bF}{{\overline F}}
\newcommand{\bxi}{{\bar\xi}}
\newcommand{\bpsi}{{\bar\psi}}
\newcommand{\disty}{\displaystyle}
\newcommand{\I}{{\rm i}}
\newcommand{\ba}{\begin{array}}
\newcommand{\ea}{\end{array}}
\newcommand{\be}{\begin{equation}}
\newcommand{\ee}{\end{equation}}
\newcommand{\bea}{\begin{eqnarray}}
\newcommand{\eea}{\end{eqnarray}}
\newcommand{\bi}{\begin{itemize}}
\newcommand{\ei}{\end{itemize}}
\newcommand{\sfrac}[2]{\mbox{$\frac{#1}{#2}$}}
\newcommand{\D}{{\rm d}}
\newcommand{\bmth}[1]{\mbox{\boldmath$#1$}}
\newcommand{\bbib}[1]{\begin{thebibliography}{#1}}
\newcommand{\ebib}{\end{thebibliography}}

\renewcommand{\thefootnote}{\fnsymbol{footnote}}
\usepackage{amscd,amsmath,amssymb}
\begin{document}
\thispagestyle{empty}
\begin{center}
{~}\\
\vspace{3cm}
{\Large\bf 2k-dimensional N=8 supersymmetric quantum mechanics}\footnote{Presented at the XI
International Conference on
Symmetry Methods in Physics, June 21-24, 2004, Prague}\\
\vspace{2cm}
{\large \bf S.~Bellucci${}^{a}$, S.~Krivonos${}^{b}$, A.~Nersessian${}^{c}$ and A.~Shcherbakov${}^{b}$ }\\
\vspace{2cm}
{\it ${}^a$INFN-Laboratori Nazionali di Frascati, C.P. 13,
00044 Frascati, Italy}\\
{\tt bellucci@lnf.infn.it} \\\vspace{0.5cm}
{\it ${}^b$ Bogoliubov  Laboratory of Theoretical Physics, JINR,
141980 Dubna, Russia}\\
{\tt krivonos, shcherb@thsun1.jinr.ru} \\ \vspace{0.5cm}
{\it ${}^c$ Yerevan State University and Yerevan Physics Institute,
Yerevan,  Armenia}\\
{\it Artsakh State University, Stepanakert, Nagorny Karabakh, Armenia}\\
{\tt nerses@lnf.infn.it}\vspace{2cm}
\end{center}

\begin{center}
{\bf Abstract}
\end{center}

We demonstrate that  two-dimensional $N=8$ supersymmetric quantum mechanics which inherits
the most interesting properties of $N=2, d=4$ SYM can be constructed
if the reduction to one dimension is performed in terms of the basic object, i.e. the $N=2, d=4$ vector multiplet.
In such a reduction only complex scalar fields from the $N=2, d=4$
vector multiplet become physical bosons in $d=1$,
while the rest of the bosonic components are reduced to auxiliary fields,
thus giving rise to the {\bf (2,\;8,\;6)} supermultiplet
in $d=1$. We construct the most general action for this supermultiplet with
all possible Fayet-Iliopoulos terms included
and explicitly demonstrate that the action possesses duality symmetry
extended to the
fermionic sector of theory. In order to deal with the second--class constraints present in the system, we introduce
the Dirac brackets for the canonical variables and find the supercharges and Hamiltonian
which form a $N=8$ super Poincar\`{e} algebra with
central charges. Finally,  we explicitly present the generalization of
two-dimensional $N=8$ supersymmetric quantum mechanics to the $2k$-dimensional case with a special K\"{a}hler geometry in the
target space.
\hfil
\newpage
\renewcommand{\thefootnote}{\arabic{footnote}}
\setcounter{footnote}0
\section{Introduction}
Among extended supersymmetric theories in diverse dimensions those which have
eight real supercharges are most interesting. Mainly, this interest is motivated by
the existence of off-shell superfield formulations. In the $N=2$, $d=4$ case the invention of
the harmonic superspace \cite{bible} and projective superspace \cite{pss} opened a way
for a detailed description of these theories. Another motivation comes from a possibility of
obtaining exact quantum results for $N=2$, $d=4$ theories in the famous Seiberg-Witten approach
\cite{SW1,SW2}. Finally, let us mention that supersymmetry severely restricts possible
target-space geometries. When the
number of supercharges exceeds eight, the target spaces are restricted to be symmetric spaces,
while beyond sixteen supercharges there is no freedom left.
Thus, the theories with eight supercharges are the
last case of theories with extended supersymmetries which have a rich geometric structure
of the target space (see e.g. \cite{VP1}).

One of the most investigated theories with eight supercharges is $N=2$, $d=4$ SYM theory. It
has been much explored and many exciting results have been obtained. The heart of the $N=2$, $d=4$ SYM theory
is formed by a vector supermultiplet, which describes spin-1 particles, accompanied by complex scalar
fields and doublets of spinor fields. The geometry of the scalar fields is restricted to be a
K\"{a}hler one \cite{KG} of special type. The restriction that the metric is defined by a
holomorphic function is crucial for the Seiberg-Witten approach. Other interesting properties of
the $N=2$, $d=4$ SYM theory are duality in the scalar sector \cite{SW1,SW2} and possibility of
spontaneous partial breaking of the $N=2$ supersymmetry by adding two types of Fayet-Iliopoulos (FI) terms \cite{SPB,IZ}.

In \cite{Z1} it has been shown that the theories with eight supercharges can be similarly formulated
in diverse dimensions still sharing the common properties. In this respect, the one-dimensional
case has a special status, because the standard reduction from the $N=2$, $d=4$ SYM to $d=1$ gives
rise to the $N=8$ supersymmetric theory with five bosons, i.e., the {\bf (5,\;8,\;3)} supermultiplet \cite{Z1,DE}.
Of course, after such a reduction almost all nice features of $N=2$ SYM mentioned above disappear.
Naturally, an obvious question arises whether it is possible to construct an $N=8$, $d=1$ theory which
\begin{itemize}
\item contains $2k$ (in the simplest case only two) bosonic fields with a special K\"{a}hler
     geometry in the target-space,
\item possesses the duality transformations, properly extended to
the fermionic sector,
 \item may be obtained by reduction from the
$N=2$, $d=4$ SYM,
\item has a proper place for FI terms.
\end{itemize}
In our recent paper \cite{BKN}  we constructed, within the Hamiltonian framework, the $N=8$ supersymmetric
mechanics which possesses the first two properties.
The goal of the present paper is to demonstrate that $N=8$ supersymmetric
quantum mechanics (SQM)
with all desired properties may indeed be constructed. Our main idea is to
perform the
reduction to one dimension in terms of the basic object -- $N=2$ vector
multiplet $\cal A$ instead of
the reduction in terms of a prepotential \cite{Z1,DE}. In this approach
only a complex scalar from
the $N=2$, $d=4$ vector multiplet becomes a physical boson in $d=1$, while
the rest of the bosonic components
are reduced to auxiliary fields. Thus, we finish with the {\bf (2,\;8,\;6)}
supermultiplet. In Subsection \ref{2L}
we construct the most general action for this supermultiplet with all possible FI terms included.
We also explicitly demonstrate that the action possesses duality symmetry extended to the
fermionic sector of the theory. In Subsection \ref{2H} we demonstrate that our SQM contains the
the second-class constraints. We then introduce the Dirac brackets for the canonical variables, and
construct the supercharges and Hamiltonian which form $N=8$ super Poincar\`{e} algebra with
central charges. Finally, in Subsection \ref{2kSQM} we explicitly present the generalization of our
two-dimensional $N=8$ SQM to a $2k$-dimensional case with a special K\"{a}hler geometry in the
target space.

\section{2k-dimensional  N=8 supersymmetric quantum mechanics}
In this section we describe a general superfield formalism of
$2k$-dimensional $N=8$ supersymmetric quantum mechanics \cite{BIKL2}. We  start with
the formulation of SQM in  $N=8$ superspace and conclude with the component
form of the Lagrangian and Hamiltonian. Due to an almost evident generalization
of the 2-dimensional case to the $2k$-dimensional one and to avoid unneeded
complications, we will describe in detail only the former case, explicitly
presenting just the final results for the $2k$-dimensional SQM.
\subsection{Two-dimensional N=8 SQM: Lagrangian}\label{2L}
The convenient point of departure is the $N=8$, $d=1$ superspace $\mathbb{R}^{(1|8)}$
$$
\mathbb{R}^{(1|8)}=(t,\theta^{ia},\vartheta^{i\alpha})\,,\qquad
\left(\theta^{ia}\right)^\dagger=\theta_{ia}\,,\qquad
\left(\vartheta^{i\alpha}\right)^\dagger=\vartheta_{i\alpha}\,,$$
where $i,\,a,\,\alpha=1,\,2$ are doublet indices of three $SU(2)$ subgroups of
the automorphism group of $N=8$ superspace\footnote{
We use the following convention for the skew-symmetric tensor $\epsilon$:
$\epsilon_{ij} \epsilon^{jk}=\delta_i^k$, $\epsilon_{12} = \epsilon^{21} =1.$}. In this superspace we define
the covariant spinor derivatives
\begin{equation}\label{sderiv} \ba{ll}
\disty
D^{ia}=\frac{\partial}{\partial\theta_{ia}}+\I\theta^{ia}\partial_t\,,\quad&
\disty\nabla^{i\alpha}=\frac{\partial}{\partial\vartheta_{ia}}+
\I\theta^{ia}\partial_t\,,\\[6pt]
\disty\left\{D^{ia},D^{jb}\right\}=2\I\epsilon^{ij}\epsilon^{ab}\partial_t\,,\quad&
\disty\left\{\nabla^{i\alpha},D^{j\beta}\right\}=
2\I\epsilon^{ij}\epsilon^{\alpha\beta}\partial_t\,,\\[4pt]
\disty\left\{D^{ia},\nabla^{j\alpha}\right\}=0\,.&\ea
\end{equation}
In order to construct a supermultiplet with two physical bosonic, eight fermionic and six
auxiliary bosonic components, i.e. the {\bf (2,\;8,\;6)} supermultiplet, we introduce, adhering to the first paper in \cite{BIKL2},
a complex $N=8$ superfield $\cZ$, $\cbZ$ subjected to the following constraints:
\be \label{constr}
\ba{lc}
\displaystyle D^{1a} \cZ = \nabla^{1\alpha} \cZ =0,\qquad D^{2a} \cbZ = \nabla^{2\alpha}\cbZ=0, & \quad (a) \\[1mm]
\displaystyle \nabla^{2\alpha}D^{2a} \cZ + \nabla^{1\alpha}D^{1a} \cbZ =\I M^{a\alpha} , & \quad (b)
\ea
\ee
where $M^{a\alpha}$ are arbitrary constants obeying the reality condition $\left(M^{a\alpha}\right)^\dagger=M_{a\alpha}$.
The constraints (\ref{constr}a) represent the twisted version of the standard chirality conditions, while
(\ref{constr}b) are recognized as modified reality constraints \cite{IZ}.
As we will
see below, the presence of these arbitrary
parameters $M^{a\alpha}$ gives rise to potential terms in the component action
and opens a possibility for a partial breaking of $N=8$ supersymmetry.
\\
The constraints \p{constr} leave the following components in the $N=8$ superfields $\cZ$, $\cbZ$:
\bea\label{components}
&&z = \cZ|,\quad \bz = \cbZ|, \quad
\psi^a =D^{2a}\cZ|,\quad\bpsi_a=-D^1_a\cbZ|,\nonumber\\
&&\xi^\alpha=\nabla^{2\alpha}\cZ|, \quad \bxi_\alpha=-\nabla^1_\alpha\cbZ|,\qquad
A=-\I D^{2a}D^2_a \cZ|,\quad {\bar A}=-\I D^{1a}D^1_a \cbZ,\nonumber\\
&&B=-\I\nabla^{2\alpha}\nabla^2_\alpha \cZ|,\quad
{\bar B}=-\I\nabla^{1\alpha}\nabla^1_\alpha \cbZ|,\qquad
Y^{a\alpha}=D^{2a}\nabla^{2\alpha}\cZ|,\\
&& {\overline Y}{}^{a\alpha} = -D^{1a}\nabla^{1\alpha}\cbZ|=Y^{a\alpha}+\I M^{a\alpha},\nonumber
\end{eqnarray}
where $|$ means restriction to $\theta^{ia}=\vartheta^{i\alpha}=0$.
The bosonic fields $A$ and $B$ are subjected, in virtue of \p{constr}, to the additional constraints
\be\label{constr1}
\frac{\partial}{\partial t} \left( A-{\bar B}\right)=0,\quad
\frac{\partial}{\partial t} \left( {\bar A}- B\right)=0.
\ee
For dealing with these constraints we have the following options:
\bi
\item to solve these constraints as
\be\label{sol1}
A=C+\frac{m}{2},\qquad B={\overline C}-\frac{\bar m}{2},
\ee
where $C$ is a new independent complex auxiliary field and $m$ is a complex constant parameter;
the resulting supermultiplet will be just {\bf (2,\;8,\;6)} one;
\item to insert the constraints \p{constr1} with Lagrangian multipliers in the proper action;
this option gives rise to a {\bf (4,\;8,\;4)} supermultiplet and will be considered in a forthcoming
paper \cite{BKNS1}.
\ei
Now one can write down the most general $N=8$ supersymmetric Lagrangian in the $N=8$ superspace\footnote{We use
the convention $\int dt d^2\theta_2 d^2\vartheta_2\equiv \sfrac{1}{16}\int dt\; D^{2a}D^2_a\;\nabla^{2\alpha}\nabla^2_\alpha$.}:
\be\label{actionsf}
\ba{ll}
\disty S=-&\disty \int dt d^2\theta_2 d^2\vartheta_2\left[ \cF\left( \cZ \right) -
  \frac{1}{2}\theta_{2a}\vartheta_{2\alpha}N^{a\alpha}\cZ -
  \frac{\I}{8}\left( {\bar n}\,\theta^a_2\theta_{2a}+ n\, \vartheta^\alpha_2\vartheta_{2\alpha}\right)\, \cZ \right]-\\[6pt]
\disty &\disty  \int dt d^2\theta_1 d^2\vartheta_1\left[ \cbF\left(\cbZ\right)+
\frac{1}{2}\theta_{1a}\vartheta_{1\alpha}N^{a\alpha}\cbZ -
 \frac{\I}{8}\left( n\, \theta_1^a\theta_{1a}+{\bar n}\, \vartheta_1^\alpha \vartheta_{1\alpha}\right)\, \cbZ \right].
\ea
\ee
Here $\cF(\cZ)$ and $\cbF(\cbZ)$ are arbitrary holomorphic functions of the superfields $\cZ$ and $\cbZ$, respectively, and two terms
with a constant real matrix parameter $N^{a\alpha}$ ($\left( N^{a\alpha}\right)^\dagger =N_{a\alpha}$) and a complex constant
parameter $n$ represent one-dimensional versions of two FI terms \cite{IZ}.

After integration over the Grassmann variables one obtains the component form of the action \p{actionsf}\footnote{All
implicit summations go from ``up-left'' to ``down-right'', e.g.,
$\psi\bpsi \equiv \psi^a\bpsi_a$, $\psi^2 \equiv \psi^a\psi_a$, $M^2\equiv M^{a\alpha}M_{a\alpha}$, etc.}:
\bea\label{actionc1}
S&=&\int\D t\,\Bigl\{
\left(F''+\bF{}''\right)\left[{\dot z}{\dot\bz}+\sfrac{\I}4\left(
\psi\dot\bpsi-\dot\psi\bpsi+\xi\dot\bxi-\dot\xi\bxi\right)\right]-\nonumber\\
&&-\sfrac{\I}{4}\left(F^{(3)}{\dot z}-\bF{}^{(3)}\dot\bz\right)
\left(\psi\bpsi+\xi\bxi\right)+\nonumber\\
&&+\sfrac{1}{16}\left[F''\left(2Y^2+AB\right)+\bF{}''
\left(2{\overline Y}{}^2+{\bar A}{\bar B}\right)-
F^{(4)}\psi^2\xi^2 -\bF{}^{(4)}\bpsi{}^2\bxi{}^2\right]-\nonumber\\
&&-\sfrac{1}{16}\left[F^{(3)}\left(\I A\xi^2+\I B\psi^2-4\psi^a
\xi^\alpha Y_{a\alpha}\right)+
\bF{}^{(3)}\left(\I{\bar A}\bxi{}^2+\I{\bar B}\bpsi{}^2+
4\bpsi{}^a\bxi{}^\alpha{\overline Y}_{a\alpha}\right)\right]+\nonumber\\
&&+\sfrac{1}{32}\,n\left(A+{\bar B}\right)+ \sfrac{1}{32}{\bar
n}\left({\bar A}+B\right)+\sfrac18\,NY+\sfrac18\,N{\overline Y}
\Bigr\}.
\eea
Here the holomorphic function $F(z)$ is defined as a bosonic limit of $\cF(\cZ)$
$$ F(z)\equiv \cF(\cZ)|\;. $$
Now, following the first option \p{sol1}, one may
express the auxiliary fields in terms of the physical ones using their equations of motion
\begin{equation}\label{auxiliary}\ba{ll}
\disty
C=\frac{\I\left(F^{(3)}\psi^2+\bF{}^{(3)}\bxi{}^2\right)+
\frac{m}{2}\left(\bF{}''-F''\right)-{\bar
n}}{F''+\bF{}''}\,,\quad&\\[9pt]
\disty{\bar C}=\frac{\I\left(F^{(3)}\xi^2+\bF{}^{(3)}\bpsi{}^2\right)-
\frac{\overline
m}{2}\left(\bF{}''-F''\right)-n}{F''+\bF{}''}\,,\quad&\\[9pt]
\disty
Y_{a\alpha}=\frac{\bF{}^{(3)}\bpsi_a\bxi_\alpha-F^{(3)}\psi_a\xi_\alpha-
\I\bF{}''M_{a\alpha}-N_{a\alpha}}{F''+\bF{}''}\,,\quad& \disty
\overline{Y}_{a\alpha}=Y_{a\alpha}+\I M_{a\alpha}\,.\ea
\end{equation}
Substituting these expressions back into Eq.\p{actionc1} we will get the action
in terms of the physical components
\be\label{actionc}
S=\int dt \left[ \cK -\cV \right],
\ee
where the kinetic $\cK$ and potential $\cV$ terms read as
\be\label{kin}
\cK= \left(F''+\bF{}''\right)\left[ {\dot
z}{\dot\bz}+\sfrac{\I}{4}\left( \psi\dot\bpsi-\dot\psi\bpsi
+\xi\dot\bxi-\dot\xi\bxi\right)\right]- \sfrac{\I}{4}\left(
F'''{\dot z}-\bF{}'''\dot\bz\right)\left( \psi\bpsi+\xi\bxi\right)
\ee
and
\begin{eqnarray}
\cV&=&\frac{1}{16}\,\biggl[\left(F^{(4)}-\frac{3F'''F'''}{F''+\bF{}''}\right)\psi^2\xi^2+
\left(\bF{}^{(4)}-\frac{3\bF{}'''\bF{}'''}{F''+\bF{}''}\right)\bpsi^2\bxi^2-\nonumber\\
&&-
\frac{F'''\bF{}'''}{F''+\bF{}''}\left(\psi^2\bpsi{}^2+\xi^2\bxi{}^2-4\psi\bpsi
\xi\bxi\right)+\nonumber\\
&&+2\,\frac{N^2-\I\left(F''-\bF{}''\right)NM +F''\bF{}''M^2}{F''+\bF{}''}+\nonumber\\
&&+4\I\,\frac{F'''\psi^a\xi^\alpha\left(\bF{}''M_{a\alpha}-\I
N_{a\alpha}\right) + \bF{}'''\bpsi^a\bxi^\alpha\left(F''
M_{a\alpha}+
\I N_{a\alpha}\right)}{F''+\bF{}''}+ \nonumber\\
&&+\frac{ n{\bar n}+\frac{1}{2}\left( m n -{\overline m}{\bar
n}\right)\left( F''-\bF{}''\right)+
  m{\overline m} F'' \bF{}''}{F''+\bF{}''}+\nonumber\\
&&+\I\,\frac{ F^{(3)}\xi^2\left(\bF{}'' m- {\bar n}\right)-
F^{(3)}\psi^2\left(\bF{}'' {\overline m}+ n\right)}{F''+\bF{}''}+\nonumber\\
&&+\I\,\frac{ \bF^{(3)}\bxi{}^2 \left(F''{\overline
m}-n\right)-\bF^{(3)}\bpsi{}^2\left( F'' m + {\bar
n}\right)}{F''+\bF{}''} \biggr].\label{pot}
\eea
The action \p{actionc} is invariant with respect to the $N=8$ supersymmetry which is realized on the
physical component fields as follows:
\begin{eqnarray}\label{susy1}
&&\delta z = \epsilon_{2a}\psi^a +\varepsilon_{2\alpha}\xi^\alpha ,\qquad
\delta\bz= -\epsilon_{1a}\bpsi{}^a -\varepsilon_{1\alpha}\bxi{}^\alpha,\nonumber\\
&&\delta\psi_a = \frac{\I}{2}\epsilon_{2a}\left( C+\frac{m}{2}\right)+\varepsilon^\alpha_2Y_{a\alpha}+2\I\epsilon_{1a}{\dot z},\\
&& \delta\xi_\alpha=\frac{\I}{2}\varepsilon_{2\alpha}\left( {\overline C}-\frac{\overline m}{2}\right)
-\epsilon^a_2 Y_{a\alpha}+2\I\varepsilon_{1\alpha}{\dot z},\nonumber
\end{eqnarray}
with $\epsilon_{ia}$, $\varepsilon_{i\alpha}$ being the parameters of two $N=4$ supersymmetries acting on
$\theta^{ia}$ and $\vartheta^{i\alpha}$, respectively, and with the auxiliary fields $C$ and $Y^{a\alpha}$ defined in \p{auxiliary}.

Using the Noether theorem one can find classical expressions for the conserved supercharges
corresponding to the supersymmetry transformations \p{susy1}
\begin{eqnarray}\label{Q}
&& Q^a_1=\left( F''+\bF{}''\right)\psi^a\dot{\bz} -\sfrac{\I}{4}\bF{}^{(3)}\bpsi{}^a \bxi{}^2+
  \sfrac{\I}{2}\left( \I\bF{}'' M^{a\alpha}+ N^{a\alpha}\right)\bxi_\alpha-
  \sfrac{1}{4}\left( m \bF{}''-{\overline n}\right)\bpsi{}^a\,, \nonumber\\
&& S^\alpha_1=\left( F''+\bF{}''\right)\xi^\alpha{\dot \bz} -\sfrac{\I}{4}\bF{}^{(3)}\bxi{}^\alpha \bpsi{}^2-
  \sfrac{\I}{2}\left( \I\bF{}'' M^{a\alpha}+ N^{a\alpha}\right)\bpsi_a+
  \sfrac{1}{4}\left( {\overline m}\bF{}''+ n\right)\bxi{}^\alpha\,,\nonumber\\
&&Q_{2a}=\left( Q^a_1 \right)^\dagger,\quad S_{2a}=\left( S^a_1 \right)^\dagger.
\end{eqnarray}

Let us note that our variant of $N=8$ SQM is a direct reduction of the
$N=2$, $d=4$ SYM.  So it is not unexpected that the metric of bosonic manifold
is restricted to be the \emph{special K\"{a}hler} one (of rigid type) (see, e.g., \cite{Fre})
\be\label{sk}
g(z,\bz)= F''(z)+\bF{}''(\bz).
\ee
Secondly, one may immediately check that the action  \p{actionc} exhibits the
famous Seiberg-Witten duality \cite{SW1}. Indeed, after passing to new variables
defined as
\begin{eqnarray}\label{duality}
&& {\tilde z}=F'(z),\quad \tilde\psi^{a}=F''\psi^{a}, \quad \tilde \bpsi_a=\bF''\bpsi_a,\quad
\tilde\xi^{\alpha}=\I F''\xi^{\alpha}, \quad \tilde \bxi_\alpha=-\I\bF''\bxi_\alpha,\nonumber\\
&& {\tilde F}{}''(\tilde z)=\frac{1}{F''(z)},\quad {\tilde N}{}^{a\alpha}=M^{a\alpha},\quad
{\tilde M}{}^{a\alpha}=-N^{a\alpha},\quad {\tilde m}={\overline n},\quad {\tilde n}={\overline m},
\end{eqnarray}
the action \p{actionc} keeps its form being rewritten in the new tilded variables.
Let us note that in the dual formulation the constants
$M^{a\alpha}$ and $m$, which appear in the constraints \p{constr} and \p{sol1},
are interchanged with the constants $N^{a\alpha}$ and $n$, which have
shown up in the FI-terms.
This is just a simplified version of the electric-magnetic duality \cite{SW1} for our
$N=8$ SQM case. Thus, our $N=8$ SQM possesses the most interesting peculiarities
of the $N=2$, $d=4$ SYM theory and can be used for a simplified analysis of some
subtle properties of its ancestor.

\subsection{Two\bmth{-}dimensional \bmth{N=8} SQM: Hamiltonian}\label{2H}
In order to find the classical Hamiltonian, we follow the standard procedure of
quantizing a system with bosonic and fermionic degrees of freedom \cite{quant}.
{}From the action \p{actionc} we  define the momenta $p$, ${\bar p}$, $\pi^{(\psi)}_a$,
$\bar\pi^{(\psi)a}$, $\pi^{(\xi)}_\alpha$, $\bar\pi{}^{(\xi)\alpha}$ conjugated to
$z$, $\bz$, $\psi^a$, $\bpsi_a$, $\xi^\alpha$ and $\bxi_\alpha$, respectively, as
\be\label{momenta}
\ba{lc}
\disty p=g\dot\bz-\frac{\I}{4}\, \partial_z g \left(\psi\bpsi+\xi\bxi\right),\quad
\bp= g{\dot z}+\frac{\I}{4}\, \bar\partial_z g \left(\psi\bpsi+\xi\bxi\right),& (a) \\[1.5mm]
\disty \pi^{(\psi)}_a=-\frac{\I}{4}\, g\bpsi_a,~
\bar\pi{}^{(\psi)a}=-\frac{\I}{4}\, g \psi^a, \quad
\pi^{(\xi)}_\alpha=-\frac{\I}{4}\,g\bxi_\alpha,~
\bar\pi{}^{(\xi)\alpha}=-\frac{\I}{4}g\, \xi^\alpha, &(b)
\ea
\ee
with the metric $g(z,\bz)$ defined in \p{sk}
and introduce the canonical Poisson brackets
\begin{eqnarray}\label{pb1}
&&\left\{ z,p \right\}=1,\quad \left\{ \bz, \bp\right\}=1, \qquad
\left\{ \psi^a ,\pi^{(\psi)}_b \right\}=-\delta^a_b,\quad
\left\{ \xi^\alpha ,\pi^{(\psi)}_\beta \right\}=-\delta^\alpha_\beta,\nonumber\\
&&
\left\{ \bpsi_a ,\bar\pi{}^{(\psi)b} \right\}=-\delta_a^b,\quad
\left\{ \bxi_\alpha ,\bar\pi{}^{(\xi)\beta} \right\}=-\delta_\alpha^\beta.
\end{eqnarray}
{}From the explicit form of the fermionic momenta (\ref{momenta} b,c) it
follows that the system possesses
the second--class constraints
\be
\ba{ll}
\displaystyle \chi^{(\psi)}_a=\pi^{(\psi)}_a+\frac{\I}{4}\,g\bpsi_a,&\quad
\displaystyle \bar\chi{}^{(\psi)a}=\bar\pi{}^{(\psi)a}+\frac{\I}4\, g\psi^a,\\[1.8mm]
\displaystyle \chi^{(\xi)}_\alpha=\pi^{(\xi)}_\alpha+\frac{\I}{4}\,g\bxi_\alpha,&\quad
\displaystyle \bar\chi{}^{(\xi)\alpha}=\bar\pi{}^{(\xi)\alpha}+\frac{\I}{4}\,g\xi^\alpha,
\ea
\ee
since
\be
\left\{ \chi^{(\psi)}_a, \bar\chi{}^{(\psi)b}\right\}=-\frac{\I}{2}\,g\delta_a^b ,\quad
\left\{ \chi^{(\xi)}_\alpha, \bar\chi{}^{(\xi)\beta}\right\}=-\frac{\I}{2}\,g\delta_\alpha^\beta .
\ee
Therefore, we should pass to the Dirac brackets defined for arbitrary  functions $\cal V$ and $\cal W$ as
\begin{eqnarray}
\left\{ {\cal V},{\cal W} \right\}_D &=&\left\{ {\cal V},{\cal W} \right\}-\left[ \left\{ {\cal V}, \chi^{(\psi)}_a\right\}
\frac{1}{\left\{ \chi^{(\psi)}_a, \bar\chi{}^{(\psi)b}\right\}} \left\{ \bar\chi{}^{(\psi)b} ,{\cal W} \right\} +\right. \nonumber\\
&& \left. \left\{ {\cal V}, \bar\chi^{(\psi)a}\right\}
\frac{1}{\left\{ \bar\chi^{(\psi)a}, \chi{}^{(\psi)}_b\right\}} \left\{ \chi{}^{(\psi)}_b ,{\cal W} \right\} +
\left( \chi^{(\psi)} \rightarrow \chi^{(\xi)}\right)\right].
\end{eqnarray}
As a result, we get the following Dirac\footnote{From now on the symbol
$\{\,,\}$ stands for the Dirac bracket.} brackets for the canonical variables:
\begin{eqnarray}\label{pb2}
&&\left\{ z,\,\hp \right\}=1,\quad \left\{ \bz, \bhp\right\}=1,\quad
 \left\{ \hp,\,\bhp\right\}=-\frac{\I}{2}
\frac{\partial_z g\, \bar\partial_z g }{g}\left( \psi\bpsi +\xi\bxi \right), \nonumber\\
&&
\left\{ \psi^a,\,\bpsi_b \right\}=-\frac{2\I}{g}\,\delta^a_b,\quad
\left\{ \xi^\alpha,\,\bxi_\beta \right\}=-\frac{2\I}{g}\,\delta^\alpha_\beta,\\
&&\left\{ \hp,\,\psi_a \right\}=\frac{\partial_z g}{g}\,\psi_a,\quad
\left\{ \hp,\,\xi_\alpha \right\}=\frac{\partial_z g}{g}\,\xi_\alpha, \nonumber\\
&& \left\{ \bhp,\,\bpsi_a \right\}=\frac{\bar\partial_z g}{g}\,\bpsi_a,\quad
\left\{ \bhp,\,\bxi_\alpha \right\}=\frac{\bar\partial_z g}{g}\,\bxi_\alpha,\nonumber
\end{eqnarray}
where the ``improved'' bosonic momenta have been defined as
\be
\hp \equiv p+\frac{\I}{4}\, \partial_z g\left(\psi\bpsi +\xi\bxi \right),\quad
\bhp \equiv \bp-\frac{\I}{4}\, \bar\partial_z g\left(\psi\bpsi +\xi\bxi \right).
\ee
Now one can check that the supercharges $Q_{ia}$, $S_{i\alpha}$ \p{Q}, being rewritten through
the momenta as
\bea\label{nQ}
 Q^a_1&=&\hp\,\psi^a -\sfrac{\I}{4}\,\bar\partial_z g\, \bpsi{}^a \bxi{}^2+
 \sfrac{\I}{2}\left(\I\bF{}'' M^{a\alpha}+ N^{a\alpha}\right)\bxi_\alpha-
 \sfrac{1}{4}\left( m \bF{}''-{\overline n}\right)\bpsi{}^a\,,\nonumber\\
 S^\alpha_1&=&\hp\,\xi^\alpha-\sfrac{\I}{4}\,\bar\partial_z g \, \bxi{}^\alpha \bpsi{}^2-
 \sfrac{\I}{2}\left(\I\bF{}'' M^{a\alpha}+ N^{a\alpha}\right)\bpsi_a+
 \sfrac{1}{4}\left({\overline m}\bF{}''+ n\right)\bxi{}^\alpha \,, \\
 Q_{2a}&=&\left( Q^a_1 \right)^\dagger\,,\quad S_{2a}=\left( S^a_1 \right)^\dagger\,,\nonumber
\eea
and the Hamiltonian
\be\label{ham}
H=g^{-1}\, \hp \bhp +{\cal V}
\ee
form the following $N=8$ superalgebra:
\bea\label{SA}
\left\{ Q_{ia},\,Q_{jb}\right\}&=&-2\I
\epsilon_{ij}\epsilon_{ab}\left( H-\sfrac{1}{16}\left( nm+{\bar
n}{\overline m}\right)\right)-
 \sfrac{1}{8}\,\epsilon_{ij}\left( N_a^\alpha M_{\alpha b}+
 N_b^\alpha M_{\alpha a}\right), \nonumber\\
\left\{ S_{i\alpha},\,S_{j\beta}\right\}&=&-2\I
\epsilon_{ij}\epsilon_{\alpha\beta}\left( H+\sfrac{1}{16}\left(
nm+{\bar n}{\overline m}\right)\right)-
 \sfrac{1}{8}\,\epsilon_{ij}\left( N_\alpha^a M_{a \beta}+
 N_\beta^a M_{a\alpha}\right), \nonumber\\
\left\{ Q_{1a},\,S_{2\alpha}\right\}&=&-m N_{a\alpha}-\I{\bar
n}M_{a\alpha}\,,\quad
\left\{Q_{2a},\,S_{1\alpha}\right\}=-{\overline m} N_{a\alpha}+\I n
M_{a\alpha}\,.
\eea
By these we complete the classical description of the two dimensional $N=8$ SQM. Before closing this subsection and
going on to generalize our SQM to the $2k$-dimensional case, let us briefly discuss the main peculiarity of the model.

Firstly, as has already been mentioned, the $N=8$ supersymmetry strictly fixes the metric of the target space to be
the \emph{special K\"{a}hler} one.

Next, the presence of the central charges in the superalgebra \p{SA}, like in the $N=4$ SQM case \cite{IKP},
is the most exciting feature of the model. The central charges appear only when the FI terms are added
(with the constants $N_{a\alpha}$ or $n$) and the
auxiliary fields contain the constant parts ($M_{a\alpha}$ or $m$). The existence of the nonzero central charges in the
superalgebra \p{SA} opens up a possibility of realizing a partial spontaneous breaking of $N=8$ supersymmetry.

Finally, it is worth noticing that the bosonic potential terms which appear in the Hamiltonian explicitly break
at least one of the $SU(2)$ automorphism groups. This is again very similar to the case of $N=4$ SQM \cite{IKP}.

\subsection{\bmth{2k-}dimensional \bmth{N=8} SQM}
\label{2kSQM}
The generalization of the $N=8$ two-dimensional SQM to the $2k$-dimensional case is straightforward.
The simplest one is the superfield generalization. The related steps are described in the following.
\bi
\item We introduce $k$ complex $N=8$ superfields $\cZ^{A}$, $\cbZ{}^{B}$
$(A,B=1,\ldots,k)$ each of them obeying the
same constraints \p{constr} with different constants $M^{A\,a\alpha}$:
\be\label{2kconstr}
\ba{lc}
\displaystyle  D^{1a} \cZ^{A} = \nabla^{1\alpha} \cZ^{A} =0,\qquad D^{2a} \cbZ{}^{A} = \nabla^{2\alpha}\cbZ{}^{A}=0, & \quad (a) \\[1mm]
\displaystyle  \nabla^{2\alpha}D^{2a} \cZ^{A} + \nabla^{1\alpha}D^{1a} \cbZ{}^{A} =\I M^{A\, a\alpha}. & \quad (b)
\ea
\ee
\item The components of each superfield can be defined as in \p{components} and $k$ different constants $m^{A}$ may
be introduced similarly to \p{sol1}
\be\label{2ksol1}
A^{A}=C^{A}+\frac{m^{A}}{2},\qquad B^{A}={\overline C}{}^{A}-\frac{{\bar m}^{A}}{2}.
\ee
\item The most general $N=8$ supersymmetric action reads
\begin{eqnarray}\label{actionsf2k}
S_{2k}&=&-\int dt d^2\theta_2 d^2\vartheta_2\left[ \cF(\cZ^{1},\ldots,\cZ^{k}) -
  \frac{1}{2}\theta_{2a}\vartheta_{2\alpha}\sum_A N^{a\alpha}_{A}\cZ^{A} - \right.\nonumber\\
&& \left. \frac{\I}{8}\sum_A \left( {\bar n}_{A}\,\theta^a_2\theta_{2a}+ n_{A}\,
\vartheta^\alpha_2\vartheta_{2\alpha}\right)\, \cZ^{A} \right]
+c.c.
\end{eqnarray}
where $\cF(\cZ^{1},\ldots,\cZ^{k})$, $\cbF(\cbZ{}^{1},\ldots,\cbZ{}^{k})$ are arbitrary holomorphic functions of
the $k$-superfields $\cZ^{A}$ and $\cbZ{}^{A}$,
respectively, and all possible FI terms with the constants $N^{a\alpha}_{A}$ and $n_{A}$ were included.
\ei
The rest of the calculations goes in the same way as it is done in the previous subsections. For completeness,
we present here the explicit structure of the Dirac brackets between the canonical variables
\begin{eqnarray}\label{2kPB}
&&\left\{ z^A,\,\hp_B \right\}=\delta^A_B,\quad
\left\{ \bz{}^A,\, \bhp_B\right\}=\delta^A_B,\nonumber\\
&&
\left\{ \hp_A,\,\bhp_B\right\}=-\frac{\I}{2}g^{EE'}\partial^3_{ACE} F\,\bar\partial^3_{BC'E'}\bF
\left( \psi^{a\,C}\bpsi^{C'}_a +\xi^{\alpha\, C}\bxi^{C'}_\alpha \right), \nonumber\\
&&
\left\{ \psi^{Aa},\,\bpsi^B_b \right\}=- 2\I g^{AB} \delta^a_b,\quad
\left\{ \xi^{A\alpha},\,\bxi^B_\beta \right\}=- 2\I g^{AB}\delta^\alpha_\beta, \nonumber\\
&& \left\{ \hp_A,\,\psi^B_a \right\}=g^{BC}\partial^3_{ACE}F\,\psi^E_a,\quad
\left\{ \hp_A,\,\xi^B_\alpha \right\}=g^{BC}\partial^3_{ACE}F\,\xi^E_\alpha, \nonumber\\
&& \left\{ \bhp_A,\,\bpsi^B_a \right\}=g^{BC}\bar\partial^3_{ACE}\bF\, \bpsi^E_a,\quad
\left\{ \bhp_A,\,\bxi^B_\alpha \right\}=g^{BC}\bar\partial^3_{ACE}\bF\, \bxi^E_\alpha,
\end{eqnarray}
where the metric $g_{AB}$ is defined as
\be\label{2kmetric}
g_{AB}=\frac{\partial^2}{\partial z^A \partial z^B} F(z^1,\ldots,z^k)+
 \frac{\partial^2}{\partial \bz^A \partial \bz^B} \bF (\bz^1,\ldots,\bz^k),\qquad
g^{AB}g_{BC}=\delta^A_C.
\ee
%and the final Hamiltonian of $2k$-dimensional $N=8$ SQM reads
Finally, the supercharges
\be\label{2kSC}
\ba{l}
Q_1^a=\hp_A\psi^{Aa}
    -\frac {\I}4 \bar\partial_{ABC}^3\bF\bpsi^{Aa}\bxi^{B\alpha}\bxi^C_\alpha
    +\frac {\I}2 \left(\I \bar\partial_{AB}^2\bF M^{Aa\alpha}
    +N_B^{a\alpha}\right)\bxi^B_\alpha\\[6pt]
\qquad    -\frac14 \left(\bar\partial_{AB}^2\bF m^A-\bar n_B\right)\bpsi^{Ba},\\[6pt]
S_1^\alpha=\hp_A\xi^{A\alpha}
    -\frac {\I}4 \bar\partial_{ABC}^3\bF\bxi^{A\alpha}\bpsi^{Ba}\bpsi^C_a
    -\frac {\I}2 \left(\I \bar\partial_{AB}^2\bF M^{Aa\alpha}
    +N_B^{a\alpha}\right)\bpsi^B_a\\[6pt]
\qquad    +\frac14 \left(\bar\partial_{AB}^2\bF m^A+\bar n_B\right)\bxi^{B\alpha}, \\
Q_{2a}=\left( Q^a_1 \right)^\dagger\,,\quad S_{2a}=\left( S^a_1 \right)^\dagger
\ea
\ee
and the Hamiltinian
\begin{eqnarray}
&&H_{2k}= g^{AB}\hp_A \bhp_B+ \nonumber\\
&&+\sfrac{1}{16}\left( \partial^4_{ABCD} F
-g^{EE'}\partial^3_{ABE}F\,\partial^3_{CDE'}F-
    2g^{EE'}\partial^3_{ACE}F\,\partial^3_{BDE'}F\right)\psi^{Aa}\psi^B_a\xi^{C\alpha}\xi^D_\alpha+\nonumber\\
&&+\sfrac{1}{16}\left( \bar\partial^4_{ABCD} \bF
-g^{EE'}\bar\partial^3_{ABE}\,\bF\bar\partial^3_{CDE'}\bF-
    2g^{EE'}\bar\partial^3_{ACE}\bF\,\bar\partial^3_{BDE'}\bF\right)\bpsi^{Aa}\bpsi^B_a\bxi^{C\alpha}\bxi^D_\alpha-\nonumber\\
&&-\sfrac{1}{16}\,g^{EE'}\partial^3_{ABE}F\bar\partial^3_{CDE'}\bF\left(
\psi^{Aa}\psi^B_a \bpsi^{Cb}\bpsi^D_b+
    \xi^{A\alpha}\xi^B_\alpha \bxi^{C\beta}\bxi^D_\beta -4 \psi^{Aa}\bpsi^C_a \xi^{B\alpha}\bxi^D_\alpha\right)+\nonumber\\
&&+\sfrac{1}{8}\,g^{AB}\left[ N^{a\alpha}_A N_{B\,a\alpha}-
       \I\left(\partial^2_{BC}F-\bar\partial^2_{BC}\bF\right) N^{a\alpha}_A M^C_{a\alpha}+
       \partial^2_{AC}F\,\bar\partial^2_{BD}\bF \,M^{C\,a\alpha}M^D_{a\alpha}\right]+ \nonumber\\
&&+\sfrac{\I}{4}\,g^{AB}\bigl[ \partial^3_{ACD} F
\psi^{Ca}\xi^{D\alpha}\left( \bar\partial^2_{BE}\bF\,
M^E_{a\alpha}+\I N_{a\alpha\, B}\right)+\nonumber\\
&&\qquad\qquad+\bar\partial^3_{ACD} \bF\, \bpsi^{Ca}\bxi^{D\alpha}
\left( \partial^2_{BE}F\, M^E_{a\alpha}-\I N_{a\alpha\, B}\right)\bigr]+\nonumber\\
&&+\sfrac{1}{16}\,g^{AB}\Bigl[ n_A {\bar n}_B+\sfrac{1}{2}\left(
n_Am^C - {\bar n}_A{\overline m}^C\right)\left(\partial^2_{BC}F-
\bar\partial^2_{BC}\bF\right)+m^C{\overline m}^D \partial^2_{AC}F\,
\bar\partial^2_{BD}\bF +\nonumber\\
&&+\I\partial^3_{ACD}F \xi^{C\alpha}\xi^D_\alpha\left(
\bar\partial^2_{BE}\bF m^E -{\bar n}_B\right)-
\I\partial^3_{ACD}F \psi^{C a}\psi^D_a
\left( \bar\partial^2_{BE}\bF {\overline m}^E +  n_B\right) + \nonumber\\
&&\I\bar\partial^3_{ACD}\bF \bxi^{C\alpha}\bxi^D_\alpha\left(
\partial^2_{BE} F {\overline m}^E -n_B\right)-
\I\bar\partial^3_{ACD}\bF \bpsi^{C a}\bpsi^D_a\left(
\partial^2_{BE} F m^E +  {\bar n}_B\right)\Bigr]
\end{eqnarray}
form the  superalgebra
$$
\left\{ Q_{ia},\,Q_{jb}\right\}=-2\I
\epsilon_{ij}\epsilon_{ab}\left( H-\sfrac{1}{16}\left( n_A m^A+{\bar
n}_A{\overline m}^A\right)\right)-
 \sfrac{1}{8}\,\epsilon_{ij}\left( N_{A\,a}^\alpha M^A_{\alpha b}+
 N_{A\, b}^\alpha M^A_{\alpha a}\right),
$$
$$
\left\{ S_{i\alpha},\,S_{j\beta}\right\}=-2\I
\epsilon_{ij}\epsilon_{\alpha\beta}\left( H+\sfrac{1}{16}\left(
n_A m^A+{\bar n}_A{\overline m}^A\right)\right)-
 \sfrac{1}{8}\,\epsilon_{ij}\left( N_{A\,\alpha}^a M^A_{a \beta}+
 N_{A\,\beta}^a M^A_{a\alpha}\right),
 $$
\be\label{SA1}
\left\{ Q_{1a},\,S_{2\alpha}\right\}=-m^A N_{A\,a\alpha}-\I{\bar
n}_A M^A_{a\alpha}\,,\quad
\left\{Q_{2a},\,S_{1\alpha}\right\}=-{\overline m}^A N_{A\,a\alpha}+\I n_A
M^A_{a\alpha}\,.
\ee
\section{Summary and conclusions}
In this paper we presented a new version of $N=8$ SQM with {\bf (2,\;8,\;6)}
components. This supermultiplet is obtained by a direct reduction from the
$N=2$, $d=4$ vector supermultiplet. We constructed the
most general action with all possible FI terms and explicitly showed that
the geometry of the target space is restricted to be the special K\"{a}hler one.
Apart from the
$N=8$ superfield formulation, we presented the component action with all auxiliary fields,
as well as with the physical fields only. As a nice feature, the constructed action possesses a
duality which acts not only in the bosonic sector, but also in the fermionic one.
We performed the Hamiltonian analysis and found the
Dirac brackets between the canonical variables. The supercharges and Hamiltonian form
a $N=8$ super Poincar\`{e} algebra with central charges. The latter are proportional
to the product of two constants --- one that comes from the FI terms, and the other that appears in the
superfield constraints (or in their solution) and coincides with the constant part of the
auxiliary fields. These constants are directly related to the appearance of the potential terms
in the Hamiltonian. Finally, we presented the extension of the $N=8$ two-dimensional SQM to the
$2k$-dimensional case.

These results should be regarded as preparatory for a more detailed study
of $2k$-dimensional SQM
with $N=8$ supersymmetry. In particular, it would be interesting to construct
the full quantum version
with some specific K\"{a}hler potential. Generally speaking, we believe that
just this version
of SQM could be rather useful for a simplified analysis of subtle problems
which appear in the $N=2$, $d=4$ SYM.
For example, one may try to fully analyse the effects of
non-anti-commutativity in superspace \cite{nac},
including modifications of the spectra, etc.

Another suitable project for a future study is to construct a one-dimensional analog of the \emph{c-map} \cite{cmap}.
It should relate the $2k$-dimensional $N=8$ SQM to the $4k$-dimensional one, with some special restriction on the
geometry of the latter. Preliminary results in this direction will appear in
a forthcoming paper \cite{BKNS1}.

Finally, due to the appearance of the central charges in the $N=8$ Poincar\`{e} superalgebra one may expect
the existence of different patterns of partial supersymmetry breaking, like in the $N=4$ SQM case \cite{IKP,P1}.

\section*{Acknowledgements}
This research was partially supported by the European
Community's Human Potential
Programme under contract HPRN-CT-2000-00131 Quantum Spacetime,
INTAS-00-00254 and INTAS 00-00262 grants, NATO Collaborative Linkage
Grant  PST.CLG.979389,
RFBR-DFG grant No 02-02-04002, grant DFG No 436 RUS 113/669, and RFBR grant
No 03-02-17440.
S.K. thanks INFN --- Laboratori Nazionali di Frascati  for the warm
hospitality extended to him during the course of this work.


\bigskip
\bbib{99}
\bibitem{bible} A.S.~Galperin, E.A.~Ivanov, S.~Kalitzin, V.I.~Ogievetsky, E.S.~Sokatchev:
Class. Quantum Grav. {\bf 1} (1984) 469;\\
A.S.~Galperin, E.A.~Ivanov, V.I.~Ogievetsky, E.S.~Sokatchev:
\emph{Harmonic Superspace}. Cambridge University Press, Cambridge 2001.
\bibitem{pss} A.~Karlhede, U.~Lindstr\"{o}m, M.~Ro\v{c}ek: Phys. Lett. B
      {\bf 147} (1984) 297;\\
U.~Lindstr\"{o}m, M.~Ro\v{c}ek: Commun. Math. Phys. {\bf 115} (1988) 21; \emph{ibid.}
 {\bf 128} (1990) 191.
\bibitem{SW1} N.~Seiberg, E.~Witten: Nucl. Phys. B {\bf 426} (1994) 19;
           Erratum-ibid. B{\bf 430} (1994) 485; {\tt hep-th/9407087}.
\bibitem{SW2} N.~Seiberg, E.~Witten: Nucl.Phys. B {\bf 431} (1994) 484;
   {\tt hep-th/9408099}.
\bibitem{VP1} A.~Van~Proeyen: \emph{``Vector multiplets in $N=2$ supersymmetry and its associated moduli spaces''},
Lectures given in the 1995 Trieste summer school in high energy physics and cosmology; {\tt hep-th/9512139}.
\bibitem{KG} G.~Sierra, P.K.~Townsend: in \emph{``Supersymmetry and Supergravity''}, ed. B.~Milewski
(World Scientific, Singapore, 1983);\\
S.J.~Gates: Nucl. Phys. B {\bf 238} (1984) 349.
\bibitem{SPB} I.~Antoniadis, H.~Partouche, T.R.~Taylor: Phys. Lett. B {\bf 372} (1996) 83;\\
S.~Ferrara, L.~Girardello, M.~Porrati: Phys. Lett. {\bf B376} (1996) 275.
\bibitem{IZ} E.~Ivanov, B.~Zupnik: Phys. Atom. Nucl. 62 (1999) 1043; Yad.Fiz. {\bf 62} (1999) 1110;
{\tt hep-th/9710236}.
\bibitem{Z1} B.~Zupnik: Nucl. Phys. B {\bf 554} (1999) 365, Erratum-ibid. B {\bf 644} (2002) 405;
{\tt hep-th/9902038}.
\bibitem{DE} D.-E.~Diaconescu, R.~Entin: Phys. Rev. D {\bf 56} (1997) 8045;
{\tt hep-th/9706059}.
\bibitem{BKN} S.~Bellucci, S.~Krivonos, A.~Nersessian: \emph{``$N=8$ supersymmetric mechanics on special K\"{a}hler
manifolds''}; {\tt hep-th/0410029}.
\bibitem{BIKL2} S.~Bellucci, E.~Ivanov, S.~Krivonos, O.~Lechtenfeld:
\emph{``ABC of $N=8$, $d=1$ supermultiplets''}; {\tt hep-th/0406015}, Nucl. Phys. B (to appear);

S.~Bellucci, E.~Ivanov, S.~Krivonos, O.~Lechtenfeld: Nucl.
Phys. B {\bf 684} (2004) 321; {\tt hep-th/0312322}.
\bibitem{BKNS1} S.~Bellucci, S.~Krivonos, A.~Nersessian, A.~Shcherbakov: in preparation.
\bibitem{Fre}P.~Fre:
%``Lectures on Special Kahler Geometry and Electric--Magnetic Duality Rotations,''
Nucl. Phys. Proc. Suppl. BC {\bf 45} (1996) 59; {\tt hep-th/9512043}.
\bibitem{quant} R.~Casalbuoni: Nuovo Cimento A {\bf 33} (1976) 389.
\bibitem{IKP} E.~Ivanov, S.~Krivonos, A.~Pashnev:
Class. Quantum. Grav. {\bf 8} (1991) 19.
\bibitem{nac} E.~Ivanov, O.~Lechtenfeld, B.~Zupnik: JHEP {\bf 0402} (2004) 012;
{\tt hep-th/0308012};\\
S.~Ferrara, E.~Sokatchev: Phys. Lett. B {\bf 579} (2004) 226; {\tt hep-th/0308021}.
\bibitem{cmap} S.~Cecotti, S.~Ferrara, L.~Girardello: Int. J. Mod. Phys. A {\bf 4} (1989) 2475.
\bibitem{P1}  E.~Donets, A.~Pashnev, J.~Juan~Rosales, M.~Tsulaia: Phys. Rev. D {\bf 61} (2000) 043512;
{\tt hep-th/9907224}.
\ebib
\end{document}